%% file: ms.tex
\DocumentMetadata{%
    pdfstandard=ua-2,
    lang=en-GB}

\documentclass[fleqn,usenatbib]{mnras}

\usepackage{newtxtext,newtxmath}

\usepackage[T1]{fontenc}

\DeclareRobustCommand{\VAN}[3]{#2}
\let\VANthebibliography\thebibliography
\def\thebibliography{\DeclareRobustCommand{\VAN}[3]{##3}\VANthebibliography}

\usepackage{graphicx}	
\usepackage{amsmath}	
\usepackage{comment}
\usepackage{soul}
\usepackage{xcolor}

\newcommand{\distance}{1.78} 
\newcommand{\alttext}[1]{}  

\usepackage{orcidlink}

\title[SiO Jet from S255N SMA 3]{A High Velocity SiO Jet from S255N SMA3}

\author[P. D. Klaassen et al.]{
P. D. Klaassen\orcidlink{0000-0001-9443-0463}$^{1}$\thanks{E-mail: pamela.klaassen@stfc.ac.uk },
M. Reiter\orcidlink{0000-0002-3887-6185}$^{2}$,
Y. Zhang\orcidlink{0000-0001-7511-0034}$^{3}$,
K. E. I Tanaka\orcidlink{0000-0002-6907-0926}$^{4}$,
and A. E. Guzman\orcidlink{0000-0003-0990-8990}$^{5}$.
\\
$^{1}$UK Astronomy Technology Centre, Royal Observatory Edinburgh, Blackford Hill, Edinburgh EH9 3HJ, UK\\
$^{2}$Department of Physics and Astronomy, Rice University, 6100 Main St, MS 108, Houston, TX 77005, USA\\
$^{3}$Department of Astronomy, Shanghai Jiao Tong University, 800 Dongchuan Rd., Minhang, Shanghai 200240, People’s Republic of China\\
$^{4}$Department of Earth and Planetary Sciences, Institute of Science Tokyo, Meguro, Tokyo, 152-8551, Japan\\
$^{5}$Joint ALMA Observatory (JAO), Alonso de Córdova 3107, Vitacura, Santiago, Chile0000-0003-0990-8990}

\date{Accepted XXX. Received YYY; in original form ZZZ}

\pubyear{\the\year{}}

\begin{document}
\label{firstpage}
\pagerange{\pageref{firstpage}--\pageref{lastpage}}
\maketitle

\begin{abstract}
We present ALMA Band 6 observations of the compact source SMA3 in the S255IR region, analysing both SiO (J = 5-4, 6-5) and CS (J = 5-4) emission to characterise the structure and kinematics of its outflow. The data reveal a highly collimated bipolar jet traced by SiO, reaching line-of-sight velocities of up to $\pm$60\,km\,s$^{-1}$ and de-projected velocities of $\pm\sim$80\,km\,s$^{-1}$, with a collimation ratio of $\sim$4.7 over a spatial extent of $\sim$5800\,au. The inferred dynamical age is 343\,yr. Position-velocity diagrams exhibit a continuous velocity structure with triangular morphology, consistent with a steady disk wind rather than an episodic or X-wind-driven flow. Excitation analysis yields a characteristic SiO temperature of $\sim$42\,K, and we derive  the jet mass, momentum, and kinetic energy which are indicative of a dense, mass-loaded molecular outflow. The emission shows clear chemical and kinematic differentiation: SiO traces the high-velocity jet spine, while CS is observed at lower velocities close to the protostar, and only reaches high velocities at larger distances. Despite forming within a clustered high-mass star-forming region, the jet exhibits a high degree of symmetry, collimation, and dynamical coherence more commonly associated with protostars forming in isolation or less crowded star-forming regions. This demonstrates that well-ordered, narrowly collimated, disk-driven jets can be launched and maintained even in the dense, competitive environments characteristic of high-mass star formation.
\end{abstract}




\section{Introduction}

Star formation is a key driver of galaxy evolution, governing the rate at which gas is converted into stars and shaping the thermodynamics and chemistry of the interstellar medium. Although stars form across a broad mass spectrum, the majority of stellar mass is assembled within clustered high-mass star-forming regions. Here,  the balance between gravity, turbulence, magnetic fields, and radiative feedback produces complex and dynamic environments full of stars with a distribution of masses drawn from an initial mass function (IMF). Bipolar jets and molecular outflows are important signposts of star formation across mass scales since, amongst other things, they emit on larger scales than the star-disk systems that form them, making them easier to identify, and they remove excess angular momentum from the collapsing system, enabling continued accretion onto the central protostar(s).

Jets and outflows trace various physical processes within these systems. Wide-angle molecular outflows primarily trace entrained ambient material, while narrow, highly collimated jets probe the immediate dynamics of gas close to the launching region. Bow-shocks in jets, in particular, exhibit strong velocity gradients and characteristic chemical signatures that reflect shock-driven excitation within the jet body \citep{Arce02}. Because their morphology, velocity structure, and chemistry are tied to the physics of accretion and protostellar disks, protostellar jets serve as powerful diagnostics of proto-stellar evolution.

A substantial theoretical and observational framework exists for understanding how protostellar jets are launched and collimated. Magneto-centrifugal models, in which magnetic fields thread the disk and extract angular momentum, provide a natural explanation for high collimation and high terminal velocities \citep{Blandford1982,Shu1994}. This formalism has lead to two prevalent types of winds: disk winds and X-winds. Modern reviews highlight how these models reproduce many of the observed properties of protostellar jets, including rotation signatures, acceleration profiles, and mass-loading rates \citep{Frank2014,Pascucci23}. Disk winds, for instance suggest mass loading over a range of radii with stratified outflows \citep[e.g.][]{Ferreira1997}. X-winds instead have a much more well defined launching radius with higher velocity flows \citep[e.g.][]{Shang2007}. 
Observational studies of Class~0 jets with ALMA such as the iconic HH~211 and HH~212 jets have revealed strikingly ordered and steady jet velocities, supporting the interpretation that these systems are powered by magneto-centrifugal disk winds \citep{Lee2008,Tobin2020}. Comparing the SMA3 jet to these well-studied systems provides critical context for evaluating whether it represents a steady, magnetically regulated jet or a more complex, time-variable flow. For both of these jets, the molecular component of the jet has been well characterised in SiO (see, for instance \cite{Jhan21,Lee17} and Table 1 of \cite{Pascucci23}).

Silicon monoxide (SiO) is a sensitive tracer of jet-driven shocks, since it is  enhanced when fast shocks sputter or shatter dust grains, liberating silicon that rapidly reacts to form SiO in the gas phase \citep{Schilke97}.  Shock models predict that the sputtering and partial destruction ($\sim 10$\%) of dust grains in fast shocks releases silicon into the gas phase, rapidly forming SiO \citep{Gusdorf08}. Observationally, SiO is consistently associated with the highest-velocity components of protostellar jets \citep{Codella2007}, while species such as HCO$^+$ and CS often trace slower-moving, entrained material \citep{SanchezMonge13}. More recent work has confirmed that SiO excitation is dominated by shock physics rather than ambient density or temperature \citep{Liu22}, making it an excellent probe of jet energetics and internal working surfaces.

Here, we present new high resolution ALMA observations of a jet system coming from one of the smaller continuum emitting regions in the S255N region (also known as G192.58-0.04) . This outer galaxy star forming region is between the evolved HII regions S255 and S257 \citep{Minier2007}, and has been well studied due to its rich chemistry and bright emission at long wavelengths. Previous mm observations have resolved three main cores \citep[SMA1, SMA2 and SMA3;][]{Cyganowski07} with strong dust continuum, numerous spectral line tracers, and both water and methanol masers \cite[see, e.g.][]{Zinchenko2012}. Maser parallax observations place SMA1 at a distance of 1.78 kpc \citep{Burns16}. These observations trace compact ionised emission characteristic of clustered high-mass star-forming environments, but they do not resolve the lower-mass protostellar components within the clump. SMA1 drives an energetic outflow previously seen in tracers like SiO and CO, while SMA2 and SMA3 were shown to be fainter, less chemically complex, and consistent with cooler, less evolved and less massive star formation \citep{Cyganowski07}.  In this paper we focus on the compact continuum source SMA3, which our higher resolution observations show drives a highly collimated bipolar jet traced in SiO. This paper presents new ALMA Band~6 observations of SMA3, providing sub-arcsecond resolution imaging of the sub-mm continuum, SiO, and CS emission. These data allow us to:
\begin{enumerate}
    \item characterize the shape and velocity structure of the SiO jet
    \item derive physical and kinematic properties of the jet, including temperature, mass, momentum, energy, and dynamical age
    \item evaluate whether the observed jet exhibits the velocity structure expected for steady-state magneto-centrifugal jets or wide-angle, entrainment-driven outflows
    
\end{enumerate}

The remainder of this paper is organized as follows. Section \ref{sec:observations} describes the ALMA observations and data reduction. Section \ref{sec:analysis} presents the analysis of the continuum and molecular line emission, including jet morphology, velocity structure, and excitation conditions. 
Section \ref{sec:discussion} discusses the implications of these results for jet launching and their relation to other protostellar jets. Section~\ref{sec:conclusions} summarizes our main conclusions.

\section{Observations}
\label{sec:observations}
\input{observations}

\begin{figure}
    \centering
    \newcommand{\continuumAlttext}{3 colour image of Spitzer emission towards the massive star forming region S225N focused in on the region observed with ALMA. There are white contours clustered around the plot showing the ALMA continuum emission, with peaks at the locations of the previously defined SMA 1,2,3 regions. Blue and red shifted flow lobes are plotted around SMA3, showing spatially distinct jet/outflow lob components on the sky which are highly elongated. The flow is at about a 45degree angle to vertical.}
    \includegraphics[width=1\linewidth,alt={\continuumAlttext}]{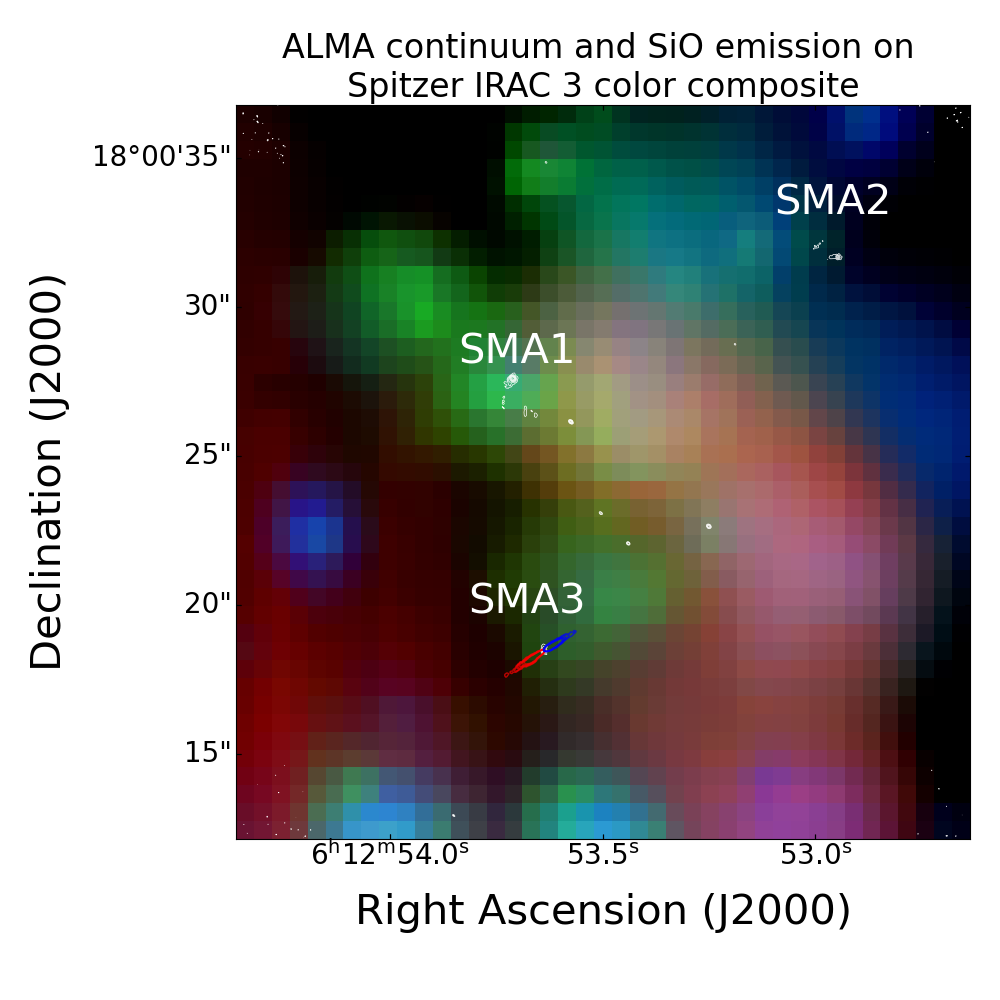}
    \caption{Spitzer 3.6, 4.5, 8.0 $\mu$m three colour (in blue, green and red, respectively) image overlayed with ALMA continuum (in white) and SiO J=5-4 emission contours for the red and blue jet lobes. The  positions of SMA1, 2, and 3 from \citet{Cyganowski07} are labelled.}
    \label{fig:continuum_context}
\end{figure}

Observations of S255N were obtained as part of project 2023.1.01346.S (PI: K. Tanaka) in Cycle 10 using two 12-m array configurations (C-4 and C-7) of band 6 observations covering two spectral setups: the first centred around 225 GHz, the other at 250 GHz. These setups allowed us to observe two transitions of SiO (the J=5-4 transition at 217.105 GHz, and the J=6-5 transition at 260.518 GHz) along with a large number of other molecular species and the dust continuum emission in both setups. The local standard of rest velocity (V$_\textrm{LSR}$) of this region is 7.8 km s$^{-1}$.  Figure \ref{fig:continuum_context} shows the ALMA continuum emission (in white contours) on the broader Spitzer IR emission in this region to put our observations into context.

Observations were taken in seven individual execution blocks, and Table \ref{tab:obs-pars} shows the  primary properties of the observations, including frequencies of the SiO spectral windows, the date, nominal configuration, and effective number of antennas in each observing set. Also shown is the on-source time per execution block, typical system temperature (T$_\textrm{sys}$) attained during the observations, baseline length ranges, and the Flux and Bandpass calibrators used for each of the execution blocks, which, in all cases was the same target. J0613+1708 was used as a phase reference source for all observations.

The data were downloaded from the ALMA archive, and the uv-plane calibrated measurement sets were restored using the CASA pipeline \citep[Version 6.6.1,][]{ALMA_pipeline}. Self calibration of the continuum was done in the pipeline, and those final data products were used here. The datasets from the two array configurations were then combined, and jointly inverted and cleaned. 
The synthesised beams and rms noise limits for all of the spectral windows used in this paper are presented in Table \ref{tab:beams}.

The object described in this paper, SMA3, was not positioned at the phase centre of our observations (which was instead SMA1), but towards the southern edge of the observed field of view. The SMA3 source centre is offset from the phase centre by 6.5" towards the southern portion of the field of view, meaning that all of the emission described in this paper falls within the half power beam width of the ALMA Band 6 observations. 

To image the spectral line data, we first used \texttt{uvcontsub} to subtract the continuum from the data in both observing configurations, combined the uv-plane data using  \texttt{concat}, and then jointly inverted into the image plane using \texttt{tclean} with interactively assigned clean boxes and natural weighting.  The same process was then repeated for the other spectral setups described in this paper.  We note that the 260 GHz spectral window was not setup with SiO 6-5 in its centre, and the highest blue-shifted emission from that line is not captured in our bandpass. While the first vibrational state of SiO 6-5 (v=1) was in our bandpass, it was not detected.

\tagpdfsetup{table/header-rows={3}}
\begin{table}
    \centering
        \caption{Beam properties for the different images used in this study}
    \begin{tabular}{lrrrr}
    \hline
     & \multicolumn{3}{c}{Synthesised Beam} & RMS\\
     & Major & Minor & Pos. Angle & noise \\
     & (arcsec)& (arcsec) & (deg) & (mJy/beam)\\
     \hline
    Continuum & 0.08 & 0.06 & 47.05 & 0.16 \\
    CS 5-4 & 0.16 & 0.12 & 30.58 & 1.04 \\
    SiO 5-4 & 0.11 & 0.08 & 135.25 & 1.34 \\
    SiO 6-5 & 0.10 & 0.07 & 40.87 & 1.61 \\
    \hline
    \end{tabular}

    \label{tab:beams}
\end{table}

\section{Analysis}
\label{sec:analysis}
In this section, we present an analysis of the continuum and molecular line emission associated with SMA3 in the S255N star forming region. We focus on both the physical structure of the protostellar core and the dynamics of its high‑velocity jet. We begin by characterizing the 250 GHz dust continuum emission to derive the core’s basic properties before turning to the morphology and kinematics of the bipolar jet traced primarily by two SiO transitions and  the complementary behaviour of CS emission along the flow. Using these spectral lines, we examine its collimation, velocity structure, dynamical age, and excitation conditions of the flow, and subsequently derive key physical quantities such as mass, momentum, and kinetic energy for each lobe. We finally use position–velocity diagrams to evaluate the velocity structures of the jet and how that relates to the terminal velocities and mass‑loading properties of the jet. 

\subsection{Continuum emission and Core properties }
\label{sec:position}

\begin{figure}
    \centering
    \newcommand{\contAlttext}{Single panel plot of a zoom in of the central few arcseconds around the protostar of insterest, SMA3. A zoom in of the ALMA continuum (shown in greyscale) and the SiO contours from Figure 1. The synthesised beam is shown in the bottom left corner. There are red contours to the bottom left of the continuum spot in the middle of the plot which show the red-shifted SiO (J=5-4) emission, and there are blue contours towards the top right of the plot which show the blue-shifted SiO (J=5-4) emission. The full extent of the flow extends beyond the plot boundary because the point of the plot is to highlight the emission in the center.}
    \includegraphics[width=\linewidth,alt={\contAlttext}]{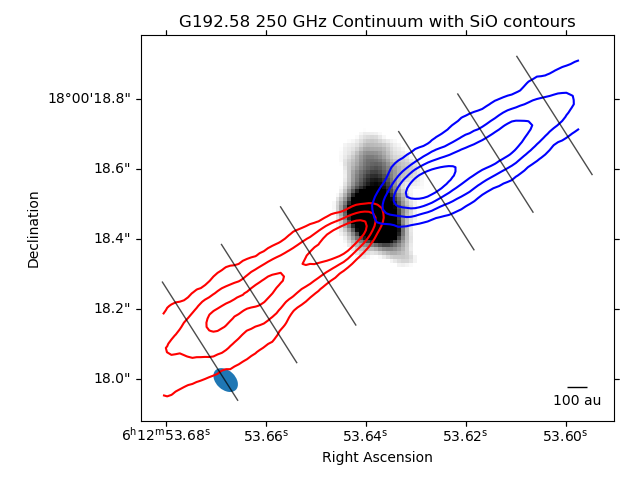}
    \caption{Continuum emission at the position of SMA3 with SiO (J=5-4) contours for the red and blue shifted jet lobes. Bars perpendicular to the jet axis are drawn at intervals described further in section \ref{sec:episodic}.}

    \label{fig:cont_contours}
\end{figure}

The 250 GHz continuum emission from SMA3 measured in our ALMA observations is shown in Figure \ref{fig:cont_contours} along with contours of the red and blue shifted SiO (J=5-4) emission. Using the \texttt{casa} task \texttt{imfit}, we fit a 2D Gaussian profile to the continuum emission peak and find a peak position of RA = 06:12:53.6384 $\pm$ 0.0002s and DEC = +18.00.18.467 $\pm$ 0.003$''$, and a resolved source size of  110.1 $\pm$ 10.3 mas $\times$ 80.9$\pm$8.9 mas at a position angle of 16$\pm$17 deg when deconvolved from the synthesised beam.

If we assume that the major and minor axes of the continuum emission are indicative of major and minor axes of a geometrically thin and circular protostellar disk, then we can derive an inclination angle for that disk. Using the values and uncertainties above, we derive an inclination of 42.7$\pm9^{\circ}$. However, since this is expected to be a very young region, it is unclear whether this assumption is valid, and we could instead be seeing a geometrically thick protostellar envelope.

The integrated intensity of the fit to the continuum emission is 18.6 $\pm$ 1.2 mJy, which we can use to derive a mass for the protostellar disk/envelope. Using a temperature of 20K for the emission surrounding the protostar, we can derive the mass of emitting dust using the following equation derived from equation 4 of \cite{Hildebrand1983}:

\begin{equation}
    M_{\textrm{dust}} = \frac{F_\nu d^2}{\kappa_\nu B_\nu (T_{\textrm{dust}})}
\end{equation}

\noindent where $F_\nu$ is the flux from the object, $d$ is the distance to it, $\kappa_\nu$ is the dust opacity, and $B_\nu(T_{\textrm{dust}})$ is the blackbody function with $\nu =$ 250 GHz and  $T_\textrm{dust}$ = 20K. Using the integrated flux calculated above, and the distance to source of 1.78 kpc, we find a dust mass of $\sim$0.007$\pm$0.005 M$_\odot$, or a total gas and dust mass of $\sim$0.7$\pm$0.05 M$_\odot$ assuming a gas to dust ratio of 100.

There are a number of chemical species emitting at and around the protostar powering this jet, including molecules like CS and CH$_3$CN, some of which are spatially resolved, most of which are not. It could be expected that these higher density tracers might show rotational signatures in the protostellar core or potentially even a protostellar disk, however their dynamics are dominated by the jet velocities, and any rotational signal cannot be distinguished from their emission profiles. Other than describing the larger scale CS emission with respect to the jet, we do not discuss the molecular emission from the core further in this work, and leave that discussion to future work.

\subsection{Jet Morphology and Velocity}
\label{subsec:jet_description}

\begin{figure*}
    \centering
    \newcommand{\SiOmomentsAlttext}{Two panel plot showing the full extent of the flow in SiO J=5-4. The left plot shows the integrated intensity plot with red and blue contours overlayed (the same ones as from the previous figure). The right plot shows the intensity weighted velocity, with blue and red scalings (via the 'jet' colourbar) showing the velocities of the dominant gas at all positions. There is a scale bar of 100au in the bottom right corner of both panels. }
    \includegraphics[width=0.45\linewidth,alt={\SiOmomentsAlttext}]{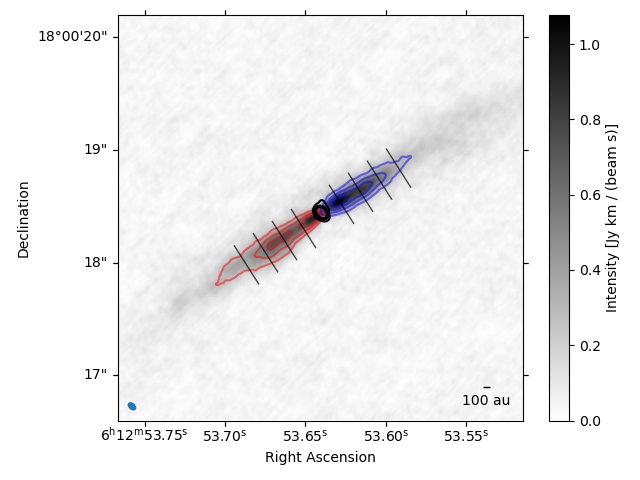}
    \includegraphics[width=0.45\linewidth,alt={\SiOmomentsAlttext}]{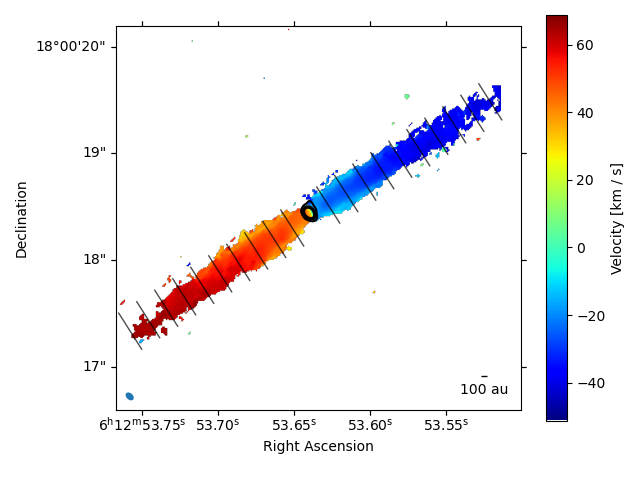}
    \caption{Moment 0 (left) and moment 1 (right) maps of the SiO 5-4. The black contours in both panels show the 250GHz continuum emission, with the blue and red contours in the left panel highlighting the integrated intensities of the red and blue emission. In the right panel, there do appear to be slight variations between the intensity weighted velocities at the edges of the blue- and red-shifted lobes when compared to their centers consistent with the gas being slowed down at the edges of the flow.}
    \label{fig:SiO5_moments}
\end{figure*}

\begin{figure*}
    \centering
    \newcommand{\SiOsixAlttext}{Two panel plot showing the full extent of the flow in SiO J=5-4. The left plot shows the integrated intensity plot with red and blue contours overlayed (the same ones as from the previous figure). The right plot shows the intensity weighted velocity, with blue and red scalings (via the 'jet' colourbar) showing the velocities of the dominant gas at all positions. There is a scale bar of 100au in the bottom right corner of both panels. }
    \includegraphics[width=0.45\linewidth,alt={\SiOsixAlttext}]{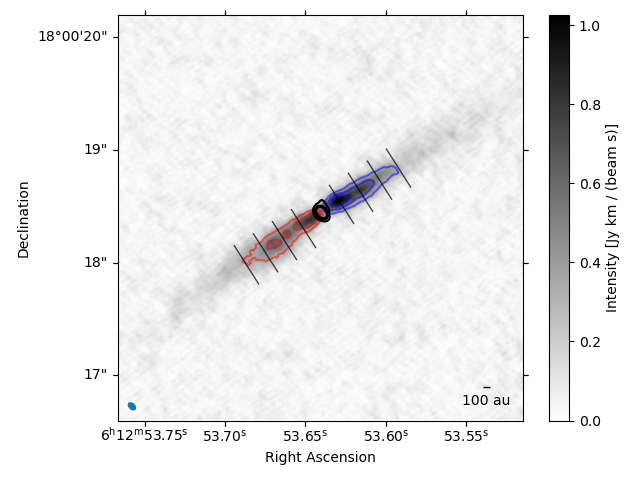}
    \includegraphics[width=0.45\linewidth,alt={\SiOsixAlttext}]{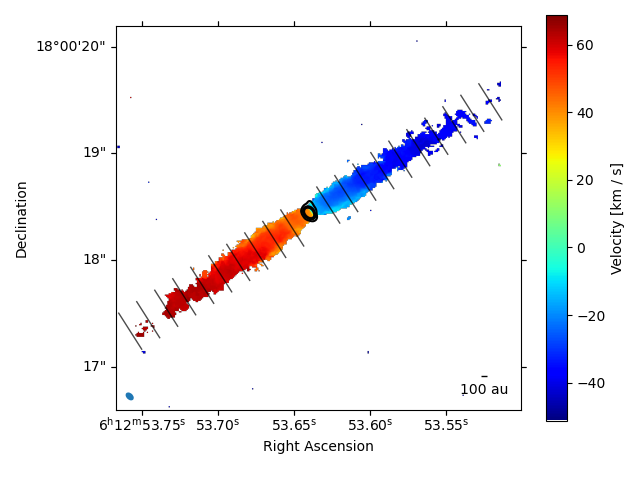}
    \caption{Moment 0 (left) and moment 1 (right) maps of the SiO J=6-5. The black contours in both panels show the 250GHz continuum emission, with the blue and red contours in the left panel highlighting the integrated intensities of the red and blue emission. In the right panel, there do appear to be slight variations between the intensity weighted velocities at the edges of the blue- and red-shifted lobes when compared to their centers consistent with the gas being slowed down at the edges of the flow. Black bars are drawn at 0.2$''$ intervals to highlight the episodic nature of the jet suggested in Section \ref{sec:episodic}.}
    \label{fig:SiO6_moments}
\end{figure*}

\begin{figure}
    \centering
    \newcommand{\SiOspectraAlttext}{Single panel plot of spectra for SiO (J-5=4) in blue, SiO (J=6-5) in orange, and CS (J=5-4) in green. The Sio spectra show little to no emission at the rest velocity of the protostar,  increasing emission at higher velocities, and a sharp drop to zero intensity at about +/- 60 km/s with respect  to the rest velocity of the target. The CS emission shows a strong peak at the source rest velocity, and a knot of emission at about -30 to -55 km/s in the blue shifted wing. Its peak roughly aligns with that of both SiO spectra on that side of the plot.}
    \includegraphics[width=0.95\linewidth,alt={\SiOspectraAlttext}]{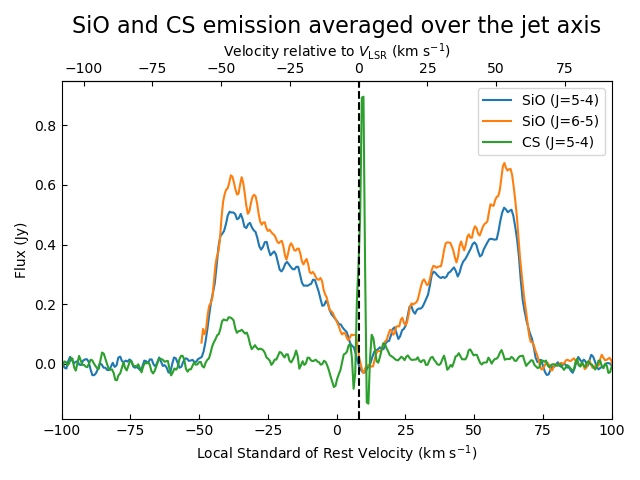}
    \caption{SiO and CS Spectra averaged over the jet. In both SiO transitions, we see that the emission increases at greater velocities, with a sharp cutoff velocity of about 60-65 km s$^{-1}$ from the source velocity of 8 km s$^{-1}$, as highlighted in the top x-axis. The SiO J=6-5 emission was at the edge of our observed spectral window, which is why there is no blue-shifted emission beyond $\sim$ -50 km s$^{-1}$. In CS (green) we see that most of the emission is at the rest velocity of the source, with an additional high velocity (blue shifted) bullet of emission that spatially corresponds to the top of the blue lobe as seen in the first moment map (right panel) of figure \ref{fig:CS_moments}. The dashed line at 7.8 km s$^{-1}$ shows the rest velocity of the target.}
    \label{fig:SiO_spectra}
\end{figure}

\begin{figure*}
    \centering
    \newcommand{\CSmomentsAlttext}{Two panel plot showing the full extent of the flow in CS J=5-4. The left plot shows the integrated intensity plot with red and blue contours overlayed. There is a lot of extended CS emission shown both along the jet direction and from the ambient material.. The right plot shows the intensity weighted velocity, with blue and red scalings (via the 'jet' colourbar) showing the velocities of the dominant gas at all positions. The high velocity CS emission appears in the top right and bottom left of the plot, with the red and blue colours appearing further out from the protostar than the SiO contours. }
    \includegraphics[width=0.95\linewidth,alt={\CSmomentsAlttext}]{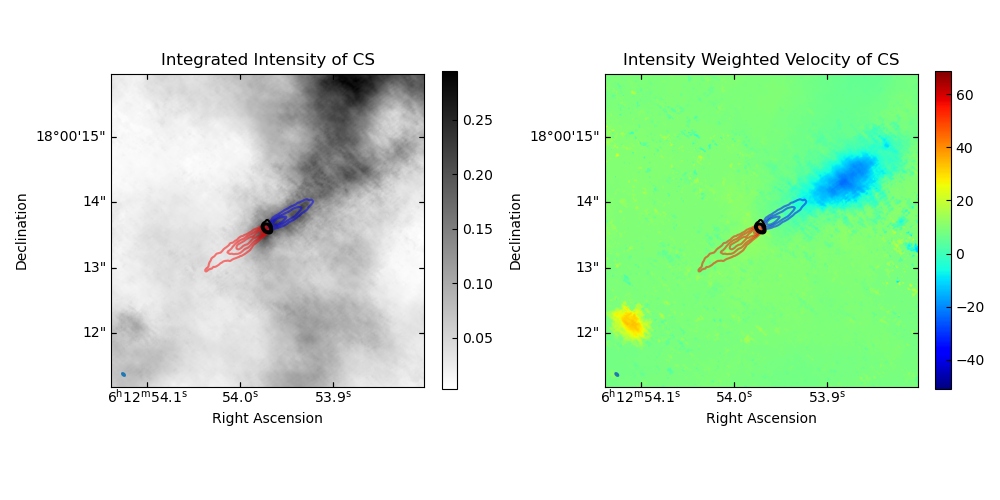}
    \caption{CS emission towards SMA3, with contours of the continuum (black) and both red and blue shifted SiO (J=5-4) integrated intensities (in red and blue contours, respectively) overplotted. \textbf{Left Panel:} Integrated intensity (moment 0) map as greyscale clipped at 3 $\sigma$, showing an abundance of emission beyond the extent of the SiO Jet. \textbf{Right Panel:} Intensity weighted velocity (moment 1) map showing the blue and red shifted CS emission beyond the extent of the SiO jet   }

    \label{fig:CS_moments}
\end{figure*}

The SiO emission from SMA 3 is highly collimated and appears as two spatially distinct velocity components on the sky.  Figure \ref{fig:SiO5_moments} shows the integrated intensity (moment zero) and intensity weighted velocity (moment one) maps of the SiO (J=5-4) emission centred on SMA3, while Figure \ref{fig:SiO6_moments} shows the same maps for the SiO (J=6-5). For both transitions, the velocity ranges (with respect to the local standard of rest, LSR, velocity) used for creating the moment maps was $\pm$ 5-60 km s$^{-1}$.  As noted in Section \ref{sec:observations}, our observations are potentially insensitive to the highest velocity blue shifted emission from the J=6-5 transition of SiO. As such, it is possible that our observations are not sensitive to the full extent of the jet in that transition given the general trend of high velocity material emitting far from the protostar. However, the blue jet lobe as seen in the J=6-5 transition in Figure \ref{fig:SiO6_moments} does appear to have a similar size as that in the J=5-4 transition in Figure \ref{fig:SiO5_moments}, suggesting any missing jet length is minimal.  We assume for the rest of this work that we have captured the blue jet emission in SiO 6=5, but do not use the length of the blue jet lobe for any calculations. The SiO spectral profiles along the jet are shown in Figure \ref{fig:SiO_spectra}, and clearly show two distinct velocity components, with little emission at the rest velocity and sharp high velocity limits for both lobes in both transitions.  The CS, also plotted in Figure \ref{fig:SiO_spectra}, shows a more characteristic Gaussian line profile centered at the source velocity, but a highly blue shifted knot of emission is also detected, and likely coincides with the blue knot of emission shown in Figure \ref{fig:CS_moments}.

The moment one maps of both species show that there is significant high velocity gas emitting at the furthest points in the outflow lobes. The intensity weighted velocity structure of the jet lobes (in both red/blue and in both transitions) show a similar layering pattern of the lowest velocity material emitting most intensely along the edges of the jet, near the protostar, with the highest velocity materials emitting in a more centrally concentrated way along the jet axis and extending further from the central protostellar system.  Deriving a jet collimation ratio from the red lobe (in both transitions of SiO), by dividing the full jet lobe extent by its largest width, we find a collimation ratio of 4.75.

As shown in Figure \ref{fig:CS_moments}, it is not only at the target position that CS emission has been detected, but along the flow as well.  The right panel of Figure \ref{fig:CS_moments} highlights the velocities of the CS gas and shows how the blue shifted emission intensifies as the SiO emission falls off further from the source. There also appears to be a redshifted knot of CS emission along the direction of the redshifted SiO lobe. It is not contiguous with the SiO emission, appearing further out from the protostar.

\subsubsection{Opening and inclination angles and their implications for velocities and jet age }
\label{sec:angle}

There are two key methods for deriving an inclination angle for a spatially resolved flow like that studied here. We can either assume that the continuum emission is a geometrically thin disk and calculate an inclination from that, or, because there is little overlap in the outflow lobes, we can use the opening angle of the flow as a proxy for a maximum inclination. We explore both options below and use their mean value for further analysis, since the former defines a lower limit and the latter an upper limit on the true inclination angle.

As discussed in section \ref{sec:position}, the continuum emission from the protostar is spatially resolved, and from the ratio of major and minor axes, we can derive an inclination angle for the system if we assume the emission is from a geometrically thin disk. Under these conditions, our deconvolved source size gives a suggested disk inclination angle of 42.7$\pm$9 deg. Since the protostar in the core is still deeply embedded and young, we cannot confirm whether any disk is geometrically thin, and so we can only use this as a lower limit on inclination angle.

\begin{figure}
    \centering
    \newcommand{\openingangleAlttext}{A close-up of the continuum emission (black contours) and SiO emission. The overall SiO emission is in the colourscale, with a single yellow contour at the 50\%  power point. Red and blue lines are drawn from the center of the continuum emission to the edges of the yellow contour, and a jet opening angle is derived from the resulting triangle.}
    \includegraphics[width=0.95\linewidth,alt={\openingangleAlttext}]{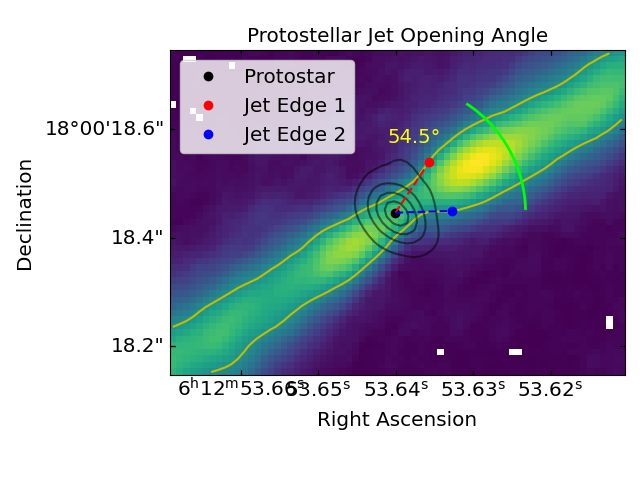}
    \caption{Opening angle of the blue lobe of the jet as derived from extending lines from the continuum peak (taken for the position of the central protostar) to the edges of the SiO integrated intensity half power. The colourscale shows the integrated intensity of the SiO (J=5-4), the black contours show the 10, 20, 30 and 40 $\sigma$ continuum emission, and the yellow contour shows the half power contour of the SiO integrated intensity. The green arc shows the opening angle projected out beyond the intensity peak after the jet becomes more linear.}
    \label{fig:opening_angle}
\end{figure}

Because the jet is highly collimated, and spatially resolved in our observations - opening to a width of 0.3$''$ or $\sim$ 500 au (compared to our < 0.1$''$ beam), we can derive an opening angle for the jet. Assuming azimuthal symmetry, we can use the plane of sky opening angle as a proxy for and upper limit on the overall opening angle. We used the  position of the protostar fitted in section \ref{sec:position} as the starting point for the jet, and the `half power' contour from the integrated intensity of SiO (J=5-4) as shown in Figure \ref{fig:opening_angle}. This method is similar to that used in \citep{Offner2011} and \citep{Dunham2024}, however they instead used the `quarter power' contour for their analysis because of better signal to noise ratios in their moment maps. From that, we traced the edges of the jet from the protostar to where the emission inflects from opening to collimated (as indicated by the red and blue dots in Figure \ref{fig:opening_angle}). From the three positions labelled in Figure \ref{fig:opening_angle}, we derived an opening angle of 54.5$^\circ$.

There does not appear to be much overlap in the red and blue emission in SiO, which suggests little inclination away from the plane of the sky. We can use the opening angle derived above therefore, as a proxy for a maximum inclination angle of the jet with respect to the line of sight under the assumption that the opening angle of the jet is equal to, or smaller than the inclination towards the line of sight.  If the inclination were lower than the opening angle, there would be spatial overlap in the emission.

The mean of these two values is 48.6$^{\circ}$. We only have formal uncertainties on the `disk' derived inclination angle, since the `flow' derived value was measured by eye. However, expanding the uncertainty on the mean to include the lower limit from the `disk' derived inclination results in roughly $48.6\pm15^\circ$, which we will use for the remainder of this work.

The radial velocities of the SiO gas peak around $\pm$ 60\,km\,s$^{-1}$ from the LSR velocity of the target in both the red and blue shifted lobes.
Given that those velocities are measured across the full length of the jet (see the PV analysis in Section \ref{sec:PV}), we can derive an dynamical age of the outflow assuming that that gas initially left the protostellar environment at that same velocity.

  Using the opening angle derived above, we can derive the total velocity of the jet from: 

\begin{equation}
    v_{\textrm{tot}} = \frac{v_\textrm{los}}{\sin(i)}
\end{equation}

\noindent where $i$ is the inclination angle , $v_\textrm{los}$ is the line of sight velocity measured in the observed emission ($\sim$ 60 km s$^{-1}$), and $v_\textrm{tot}$ is the total velocity of the SiO gas. Using this equation, we find $v_\textrm{tot} \simeq 80\pm18$ km s$^{-1}$. This derived, de‑projected, jet velocity lower limit falls within the predicted range for magneto‑centrifugal disk winds \cite{Pascucci23}, supporting a disk‑wind origin for the SMA3 jet.

Using the same inclination angle to de-project the plane-of-sky jet length,  the average measured plane of sky SiO jet lobe length of 2.15$''$ (or 3830 au at a distance of 1.78 kpc) translates to a jet length of 5800$\pm$1700 au.

Using these two de-projected values together, we can derive an upper limit to the kinematic age for the outflow (the time it would take 80 km s$^{-1}$ gas to travel 5800 au). We derive a kinematic age for the SiO outflow 343$\pm$181 yr.

Of course, the flow itself is much older and longer than this, as shown by the collimated and high velocity CS emission seen in Figure \ref{fig:CS_moments}. The enhancement of \textit{high‑velocity} CS emission at larger separation from the protostar than SiO agrees with the expected stratification in shock‑driven flows, where SiO traces the immediate high‑velocity jet while CS reflects denser entrained post‑shock gas \cite{SanchezMonge13}. 

We can also apply the same age analysis to the CS emission, which has a peak line of sight velocity of 50 km s$^{-1}$ in both the red and blue lobes. The far ends of the CS emission are also highly symmetric, with emission reaching 2.7$''$ from the protostar as shown in Figure \ref{fig:CS_moments}. Using the same inclination angle as the SiO jet, we calculate an outflow length of 7270$\pm$2160 au and derive a kinematic age of 520$\pm$270 yr for the high velocity CS emission, with the caveat that any internal shocks may be decreasing the overall flow velocity.

\subsection{Position-velocity diagrams, terminal velocities and episodic emission}
\label{sec:PV}

 Position-velocity (PV) diagrams provide a powerful diagnostic of jet launching, mass loading, and shock propagation in protostellar systems. Because they encode both spatial and kinematic structure, PV diagrams preserve the imprint of the jet–launching region, including velocity layering, internal working surfaces, and ambient–jet interactions. In the earliest low-mass protostellar stages, jets are frequently molecular, rich in CO, SiO, SO, CS, and related tracers \citep{Lee20Review,Dutta25}, reflecting dense environments where cooling and shock processing are efficient.

\begin{figure*}
    \centering
    \newcommand{\PVdiagAlttext}{Two panels of position-velocity diagrams showing velocity on the y-axis and position along the flow on the x-axis. The mission is in the top left and bottom right of each panel, showing a roughly triangular shape in each lobe. dashed lines show the rest velocity and continuum peak positions while dotted vertical lines show the spacings of suspected emission knots.}
    \includegraphics[width=1\linewidth,alt={\PVdiagAlttext}]{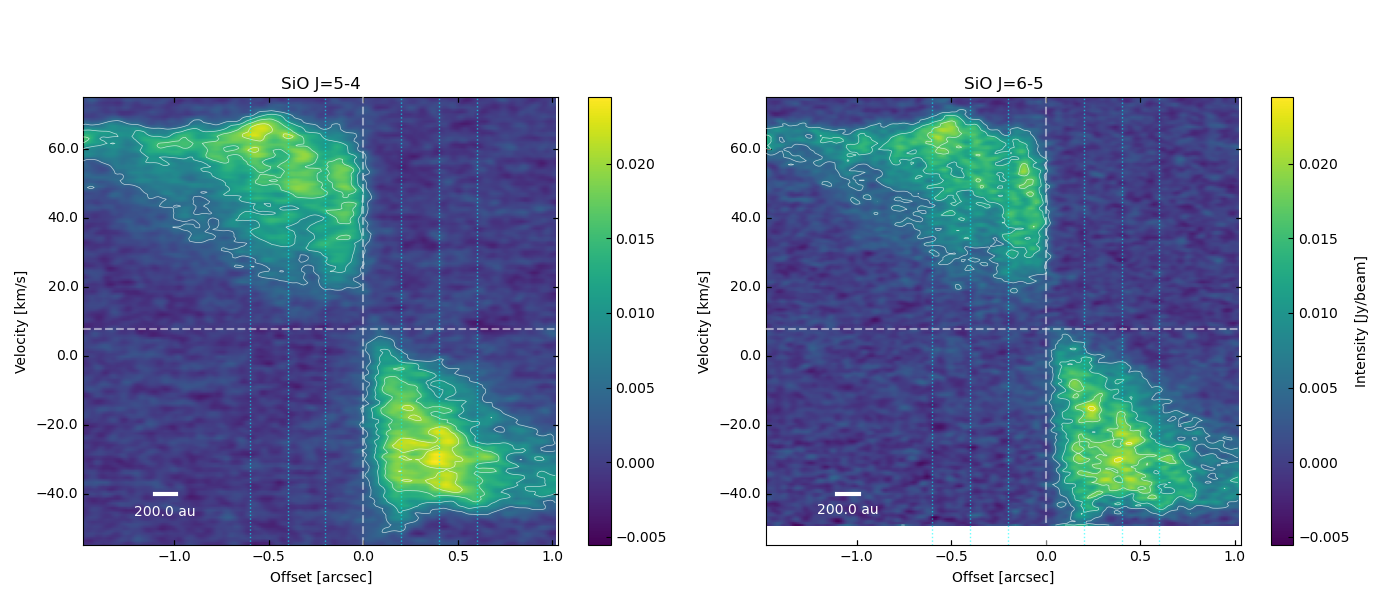}
    \caption{PV diagrams of the SiO J=5-4 and J=6-5 emission showing that the highest velocity gas is furthest from the target. }
    \label{fig:PV_diagrams}
\end{figure*}

Because PV diagrams show how the gas is moving as a function of position along the jet, they provide information on the driving mechanisms behind the outflowing material \citep[see, e.g. Figure 1 of ][]{Arce02}. In Figure \ref{fig:PV_diagrams}, we present PV diagrams for the SiO emission (J=5-4 and J=6-5, respectively in the left and right panels) along the jet axis, where the LSR velocity (7.8 km s$^{-1}$) and position of the driving source are highlighted by dashed lines.

When accounting for the V$_\textrm{LSR}$ of the source, the highest line-of-sight velocities reached in the emission are $\pm \sim$60~km\,s$^{-1}$. 
Both jet lobes show triangular morphologies in the PV diagrams, showing that there is high velocity gas at all distances, but no low velocity gas at large distance. Importantly, and consistently within the first moment maps in Figures \ref{fig:SiO5_moments} and \ref{fig:SiO6_moments}, the base of the flow exhibits a clear velocity spread rather than a single narrow high-velocity spike. This morphology matches the expected PV signature of a disk wind launched over a range of radii, rather than an X-wind confined to a narrow launching annulus. 

The PV structure with its layered velocity distribution close to the source and  moderate inclination-corrected terminal speeds of $\sim$80~km\,s$^{-1}$, is therefore most consistent with a \emph{magneto-centrifugal disk wind}. The presence of strong SiO and CS emission and the shape of the emission in the PV diagrams indicate substantial shock processing consistent with shocked dense gas \citep{Gusdorf08,Anderl2013,Flower99}. Together, these PV signatures support the interpretation of a  shock-rich, disk-wind-driven molecular jet, in agreement with the broader theoretical, numerical, and observational framework for jets from young low-mass protostars.

The contours in Figure \ref{fig:PV_diagrams} start at 3 sigma, and increase in increments of 3 sigma. In both diagrames thare are vertical stripes of emission peaks. They are consistent between the two transitions and appear at roughly 0.2$''$ intervals in both the red- and blue-shifted parts of the diagrams. Those intervals are shown as vertical dashed lines the figure.

\begin{figure}
    \centering
    \newcommand{\tempAlttext}{Image of the derived temperatures along the jet. Bars perpendicular to the jet are shown at the suggested episodic intervals, with temperature peaks coinciding with some of them.}
    \includegraphics[width=0.95\linewidth,alt={\tempAlttext}]{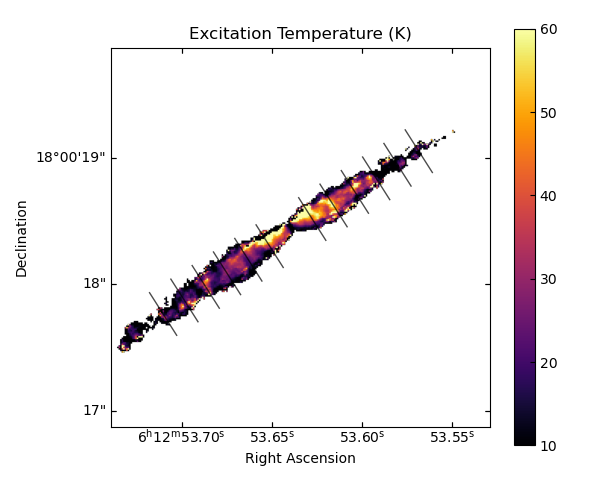}
    \caption{Excitation temperature of the SiO gas as derived from the column density ratio between the SiO J=6-5 and J=5-4 transitions. While peak temperatures are in excess of 200 K in some pixels, the colourscale here is truncated at 60 K to highlight temperature variations  across the jet.}
    \alttext{Single panel plot showing the derived temperature at all positions in the jet. The highest temperatures appear to be on the orth rim of the jet, however there are also some knots of emission along the blue lobe.}
    \label{fig:T_ex}
\end{figure}

They suggest some kind of episodic ejection mechanism is at play in the flow launching region. Equivalent 0.2$''$ spacing bars are plotted in Figures \ref{fig:SiO5_moments}, \ref{fig:SiO6_moments}, and \ref{fig:T_ex}, with the former moment maps not showing much episodic variation in the emission, but the latter temperature map (as explained in more detail in Section \ref{subsec:excitation}), showing variations on those size scales.

\subsection{Excitation Conditions Along the Jet }
\label{subsec:excitation}

With observations of two rotational transitions of SiO, we derived an excitation temperature for the jet emission using a two-level rotation diagram analysis. Using the assumption that the overall gas population is in local thermodynamic equilibrium (LTE), and that the emission is optically thin, the column density in a given energy level can be related to the overall column density of that species via:

\begin{equation}
    \frac{N_u}{N} = \frac{g_u \,\textrm{e}^{-E_u/kT_{ex}}}{Z(T_{ex})}.
\label{eqn:part_func}
\end{equation}

\noindent Here, $N_u$ is the column density in the upper level, $N$ is the total column density for that species, $g_u$ and $E_u$ are the statistical weight and energy of the upper level (respectively), $k$ is the Boltzmann constant, $T_{ex}$ is the excitation temperature and $Z(T_{ex})$ is the partition function at the excitation temperature. The data in the spectral cubes are presented with units of Jy/beam, and because the equations for deriving physical properties require brightness temperatures, the first conversion was from units of flux density (Jy/beam) to brightness temperature (K). We did this conversion using $T_b = \left( \frac{S_\nu}{\Omega} \right) \cdot \frac{c^2}{2k \nu^2}$ where $\Omega$ is the beam solid angle, $\nu$ is the frequency of the transition, $k$ is Boltzmann's constant, and $c$ is the speed of light. We used the \texttt{common\_beam} method in the \texttt{radio\_beam} package to derive $\Omega$ directly from the headers of the input \texttt{FITS} files, and as presented in Table \ref{tab:beams}.

Taking the logarithm of both sides of equation \ref{eqn:part_func} gives:

\begin{equation}
    \ln\left(\frac{N_u}{g_u}\right) = -\frac{E_u}{k T_{\text{ex}}} + \ln\left(\frac{N}{Z(T_{\text{ex}})}\right)
\end{equation}

\noindent where $1/kT_{ex}$ is the slope of the line of best fit and the total column density (divided by the partition function) is the y-intercept.  The calculation of $N_u$ relies on knowing the integrated intensity of the line, and its Einstein A coefficient ($A_{u\ell}$):

\begin{equation}
    N_u = \frac{8\pi k \nu^2}{hc^3A_{u\ell}}\int T_{mb}dv
    \label{eqn:SiO_uppercol}
\end{equation}

The calculated upper level column densities for the two transitions can be used in equation \ref{eqn:part_func} to derive the slope of the line of best fit, and from that, the excitation temperature of the SiO gas via:

\begin{eqnarray}
    \ln\left(\frac{N_a}{g_a}\right) - \ln\left(\frac{N_b}{g_b}\right)
    & = &
    - \frac{E_a-E_b}{kT_\text{ex}}\\
    \label{eqn:T_ex}
    T_{\text{ex}} &=& \frac{E_a - E_b}{k \left[ \ln\left(\frac{N_b / g_b}{N_a / g_a}\right) \right]}
\end{eqnarray}

\noindent where $a$ and $b$ represent the derived values and higher energy state constants for the J=6-5 and J=5-4 transitions, respectively. 

We calculated $\text{ln}(N_u/g_u)$  in both transitions and subsequently derived the excitation temperature on a pixel-by-pixel basis because our maps were created on the same pixel grid. We masked the data cubes for both SiO transitions at 5 sigma, and then created integrated intensity maps of the emission by summing over the channels with signal. Then, once $\text{ln}(N_u/g_u)$ was calculated for that pixel, T$_\textrm{ex}$ was derived.

Near the target, on the northern edge of the jet, the derived temperature is quite high (in excess of 200\,K), as is the uncertainty on the estimate in that region. Therefore, we take a mean temperature of the gas from the core of the two outflow lobes, excluding this region. We find a mean excitation temperature of 42\,K. We use this temperature in our kinematic calculations below.  

\subsubsection{Episodic Ejection}
\label{sec:episodic}

The episodic emission first seen in the PV diagrams of Figure \ref{fig:PV_diagrams} have led to short bars perpendicular to the jet being plotted on the temperature map at 0.2$''$ intervals based on the spacings highlighted by the dotted lines in the PV diagrams.  Especially evident in the blue-shifted lobe (top-right of the figure) are temperature enhancements in the SiO gas periodically on these intervals, with gas temperatures reaching an excess of 60K. This lends further credence to there being something pushing hotter and possibly denser material into the jet on a periodic timescale.

The periodic spacing of $\sim0.2''$ corresponds to a de-projected knot separation of $\sim540$ au and an ejection timescale of $\sim32$ yr, when using the derived jet velocity of 80 km\,s$^{-1}$. If this periodicity reflects a dynamical timescale in a Keplerian disk, this implies a radius of $\sim9$ au given the derived mass of 0.7\,M$_\odot$. This radius is significantly larger than the expected launching region of the high-velocity jet if the 80 km\,s$^{-1}$ material is being launched at its escape radius ($< 1$ au),  and therefore more likely traces the origin of variability within the disk. Several processes may produce quasi-periodic modulation of the accretion and ejection rates, including long-lived disk substructures such as pressure maxima or vortices \citep{Johansen2007,Lyra2013}, episodic accretion driven by gravitational instabilities \citep{Vorobyov2005,Vorobyov2015}, cyclic behaviour associated with dead-zone boundaries \citep{Armitage2001,Zhu2010}, or periodic perturbations induced by an embedded companion \citep{Reipurth2000,Raga2017}.  Episodic variability in the mass-loading of the wind may arise through such disk instabilities and  accretion processes, which naturally generate time-variable ejection \citep{Romanova2009}.

\subsection{Derived Flow Dynamics }
\label{subsec:flow_props}

The observed SiO spectral line emission can be used to determine the mass, momentum, and energy in the jet lobes.   With the intensity expressed as a temperature, and the data masked at 3 sigma, the column density in each velocity channel can be derived using the intensity of the line. We did that calculation \textit{per velocity channel} in each of the SiO data cubes, and summed the results to derive the full column density of the emission in each lobe for the emission in each transition. This per-channel method allows for more detailed momentum calculations.

Using the equations and methodology shown in  \cite{Lopez09}, we derive the kinematic properties of our jet. 

The upper level column densities derived above in equation \ref{eqn:SiO_uppercol} where used to derive the overall column density of the SiO gas using the partition function.  Using an SiO abundance of $10^{-7}$ with respect to molecular hydrogen, appropriate for the shocked regions \citep[see, eg][]{Schilke97}, this SiO column density was then converted to an average molecular column density per velocity channel using this conversion factor.  The column densities in each channel were then summed to derive the overall column density of the material in the jet lobe.   The SiO and overall molecular column densities calculated by summing over all velocity channels for each jet lobe are presented in Table \ref{tab:kinematics}. The uncertainties used in these calculations come from the rms noise in the spectra, and half the velocity resolution of the observations, which are propogated through the calculations in quadrature.

Given the distance to the target (\distance\, kpc) we can quantify the linear area of each pixel and with the derived column density in each pixel, we calculate the emitting mass using:

\begin{equation}
    M_\text{pixel} = N_{\text{H}_2} \cdot A_{\text{pixel}} \cdot \mu_{\text{mol}} \cdot m_p
\end{equation}
\label{eqn:mass}

\noindent where $A_{\text{pixel}}$ is the pixel area in cm$^2$, $\mu_{\text{mol}}$ is the mean molecular weight \cite[$\mu_{\text{mol}}$ = 2.8,][]{Kauffmann08}, and $m_p$ is the proton mass. The total mass is then calculated by summing the masses derived per pixel. As with column density, the mass per channel was calculated and the values presented in Table \ref{tab:kinematics} show the summed mass over the entire outflow lobe for the given SiO transition. Since, in the optically thin regime, this calculation is a summation of particles along the line of sight, no correction for inclination is required.

The momentum in the jet was calculated via:

\begin{equation}
    P = \sum_v M \cdot (v - v_{\text{LSR}})
\end{equation}

\noindent where $M$ is the mass in the velocity channel, and ($v$ - $v_{\text{LSR}}$) is the velocity of the gas in that channel corrected for the LSR velocity of the target (7.8 km s$^{-1}$) and inclination angle. The values presented in Table \ref{tab:kinematics} were summed separately over the velocity ranges for the two lobes. Similarly, the kinetic energy in each channel was calculated using:

\begin{equation}
    E_\text{k} = \sum_v \frac{1}{2} M \cdot (v - v_{\text{LSR}})^2
\end{equation}

\noindent where the sum is performed per-lobe. In these velocity dependent derivations, we have explicitly applied the inclination correction in each channel through the calculation. With the jet age derived in Section \ref{sec:angle}, we can further derive mass loss ($\dot{\textrm{M}}$), outflow force ($\dot{\textrm{P}}$) and power ($\dot{\textrm{E}}$) by dividing these values by the outflow age. These further properties are also shown in Table \ref{tab:kinematics}.

We can compare these values to those derived for flows in Ophiucus \citep[e.g.][]{Nienke13} and Serpens \citep[e.g.][]{Plunkett2015}, which themselves have a wide range in flow force values ($10^{-3}$-10$^{-7}$ M$_\odot$ km s$^{-1}$ yr$^{-1}$) . We find our derived flow kinematics are consistent with the envelopes set by these studies.

\begin{table*}
\begin{tabular}{|c|c|c|c|c|c|c|c|c|}
\hline
Lobe & N[SiO] & N[H$_2$]  & Mass  & Momentum  & Energy  & Mass {flow} Rate  & Force  & Power  \\
Colour & ($10^{16}$ cm$^{-2}$) & ($10^{23}$ cm$^{-2}$) &  ($10^{-3}$ M$_\odot$) &  ($10^{-1}$ M$_\odot$ km/s) & ($10^{43}$ erg)&($10^{-5}$ M$_\odot$/yr) &  ($10^{-4}$ M$_\odot$ km/s/yr) & ($10^{40}$ erg/yr) \\
\hline
Red & 6.34$\pm$0.02 & 6.35$\pm$0.02 & 2.183$\pm$0.008 & 1.111$\pm$0.004 & 6.11$\pm$0.03 &0.6$\pm$0.3 & 3$\pm$2 & 18$\pm$9 \\
Blue & 6.32$\pm$0.02 & 6.32$\pm$0.02 & 2.129$\pm$0.008 & -0.899$\pm$0.004 & 4.23$\pm$0.02 &0.6$\pm$0.3 & 3$\pm$1 & 12$\pm$7 \\
\hline 
\end{tabular}

\caption{Jet kinematics derived from the calculations presented in Section \ref{subsec:flow_props}. The column density of the gas component was derived from the average SiO column density assuming an abundance ratio of $10^{-7}$, and all quantities first calculated per velocity channel, then summed for the overall value presented here. The mass loss rate, force and power were derived using a jet age of 343 $\pm$181 yr, as calculated in Section \ref{sec:angle}}
\label{tab:kinematics}
\end{table*}

\section{Discussion}
\label{sec:discussion}

The SMA3 jet exhibits a high level of collimation, symmetry, and velocity coherence with a collimation ratio of $\sim 4.7$, comparable to L1551 \citep{Gerald2006} and the flows from NGC1333 IRAS4A and IRAS2A \citep{Plunkett2013}. Despite being embedded within the more complex, clustered environment of a high-mass star-forming clump, SMA3 nonetheless produces a jet that displays the morphological simplicity and dynamical regularity commonly associated with isolated low-mass protostars (i.e. BHR~71 and similar systems). The collimated, axis-aligned morphology of SMA3 suggests that its launching region is confined and stable over the dynamical timescales probed here (less than 1000 yr). This provides valuable evidence that compact, disk-anchored jets can arise even within dense and turbulent environments where multiplicity, feedback, and radiative heating could otherwise disrupt the collimation process \citep[e.g.][]{Frank2014,Qiu2011,Seifried2012,Peters2014}.

Disk winds, launched over a broad radial extent, naturally produce an ``onion-like'' velocity stratification across the jet: high-velocity components arise from inner radii, while slower material originates farther out \citep{Lee20Review,Frank2014}. Consequently, PV diagrams of disk winds show a velocity spread close to the source rather than a single, narrow, high-velocity spine. In contrast, X-wind models predict that the jet originates from a compact region near the disk truncation radius, producing a sharply peaked velocity component with minimal near-base dispersion \citep{Shu1994}. In molecular jets, cooling-driven shocks and internal working surfaces further reshape the PV structure, as demonstrated by numerical simulations of molecular jets at injection velocities of $\sim$50--100~km\,s$^{-1}$ \citep{Moraghan06,Lee01}. These internal shocks appear as tilted or curved arcs in PV space, with gradients reflecting the bow-shock geometry and the interaction between the jet and its environment \citep{Arce02}. The strong SiO and CS emission observed in many protostellar jets originates from grain sputtering and shattering in C- and J-type shocks \citep{Schilke97,Gusdorf08,Guillet2009,Anderl2013,Walmsley06,Flower99,VanLoo13}, and the resulting shock-induced chemistry often enhances the brightness and velocity width of these PV features. Observations of SiO jets such as HH\,211 and HH\,212 show these signatures clearly \citep{Codella2007,Lee2008,Yoshida21}, and environmental excitation differences can further modulate the PV appearance \citep{SanchezMonge13,Penaloza18,Liu22}.

The PV diagrams of SiO emission provide some of the strongest indicators that the SMA3 jet is driven by a steady, rather than episodic, ejection mechanism. Both SiO transitions show a two-sided triangular PV shape, with diagonal boundaries. This structure is characteristic of  a jet with nearly constant terminal velocity along the flow axis with lower velocities arising from interactions with the surrounding medium, and closely resembles the steady-jet models presented by \cite{Lee01}.

In contrast, pulsed or episodic jets would produce multiple PV ridges, arcs, or bow-shock-like loops corresponding to successive ejection bursts. No such sub-structure is observed in the SMA3 PV diagrams. Instead, the simple, linear velocity gradients and symmetry between the red and blue lobes reinforce the interpretation that the SMA3 jet is dynamically stable on $\lesssim$ 1000 year timescale. 

\subsection{Chemical Stratification: The Relationship Between SiO and CS Emission}

A key feature of the SMA3 jet is the distinct spatial and velocity separation between SiO and CS emission. The SiO traces the highest-velocity material and is concentrated along the spine of each jet lobe, consistent with its formation in fast shocks where dust grains are sputtered or shattered. The CS emission, by contrast, becomes more apparent in the first moment maps (i.e. has high velocity emission)  further down the flow, only after the SiO intensity declines as shown in Figure \ref{fig:CS_moments}.

This offset is consistent with expectations for chemically layered shocks: SiO is produced promptly in high-velocity, grain-processing regions, while CS forms or survives preferentially in the denser, cooler post-shock gas. The fact that CS is less prevalent in the innermost, highest-velocity regions suggests that it may be destroyed there on relatively short timescales, potentially fragmenting into species such as OCS, SO, or HCS. The appearance of red-shifted CS knots further out in the lobe supports the idea that CS traces the entrained or re-processed material once the gas has slowed, cooled, and re-established molecular abundances. That the moment 1 map in Figure \ref{fig:CS_moments} shows an overall constant, and zero velocity relative to the LSR, velocity field except for at the ends of the jet suggests that there is little to no feeding of the source from any surrounding streamer or filament. Together, the SiO and CS distributions reveal a stratified jet in which different chemical zones trace distinct dynamical components of the flow.

\subsection{Implications for Jet Launching and Disk--Jet Coupling}

While the inferred periodicity discussed in Section \ref{sec:episodic} corresponds to dynamical timescale at this radius of $\sim9$ au, this should not be interpreted as a jet launching radius. Many launching models generally place the origin of the high-velocity jet within the inner $\lesssim1$ au of the disk where escape velocities are high, with X-wind models confining the launch region to the vicinity of the magnetospheric truncation radius and disk-wind models invoking a range of footpoints extending outward through the inner disk \citep{Shu1994,Ferreira2006,Frank2014}. For a 0.7\,M$_\odot$ protostar, the observed jet velocity of $\sim80$ km\,s$^{-1}$ is consistent with launching from radii of order 0.5--1 au from magnetocentrifugal acceleration models. We favour an interpretation in which the observed knots in the wind launched from within 1 au arise from some form of modulation process or structure further out in the disk (at the $\sim9$ au radius). 

Taken together, the jet morphology, PV structure,  and chemical stratification converge on a consistent picture in which the SMA3 jet is launched by a narrow, magneto-centrifugal disk wind rather than a wide-angle outflow or turbulent burst. The observed opening angle, lack of lobe overlap, and high degree of symmetry all indicate a relatively confined and stable launching region.

Moreover, the presence of dusty material in the jet, inferred from the presence of SiO, suggests that the launching region may capable of lifting dust grains above the disk midplane. However, given the timescales of the jet, this could also be simply the product of SiO being lifted into the jet.   Recent high-resolution interferometric studies of young disks have shown that dusty disk winds can arise from within sub-au radii, providing a natural mechanism for injecting dust into the base of the jet before grain destruction occurs. Such winds offer an elegant link between disk structure, jet chemistry, and the dynamic behaviour observed on larger scales.

Overall, the SMA3 jet appears to be a compact and well-regulated flow powered by a disk wind with a characteristic terminal velocity of $\sim 74~\mathrm{km~s^{-1}}$. Its properties align closely with theoretical expectations for steady magneto-centrifugal jets and reinforce the idea that even deeply embedded protostars can sustain highly collimated and dynamically ordered outflows.

\section{Conclusions}
\label{sec:conclusions}

In this paper, we have presented new ALMA Band~6 observations of S255N SMA3 that resolve, in detail, the morphology, velocity structure, excitation, and chemistry of a highly collimated SiO jet. The combined continuum and spectral–line analysis reveals a jet with a de-projected terminal velocity of $\sim 80$\,km\,s$^{-1}$, derived from line–of–sight velocities of $\pm 60$\,km\,s$^{-1}$ and an inclination angle of $48.6^\circ$ taken as the average of jet opening angle and disk inclination angle arguments. The SiO lobes extend $1.65''$ on the sky, corresponding to a de-projected physical length of $\sim 5800\pm1700$\,au. Together, these measurements imply a kinematic jet age of $\sim 343\pm181$\,yr, demonstrating that SMA3 is driving a stable and dynamically coherent molecular outflow. The high collimation ratio ($\sim 4.75$), strong symmetry between the red and blue lobes, and the absence of internal substructure in the PV diagrams all point toward a directly launched flow rather than an entrained one.

Our excitation analysis yields a characteristic SiO excitation temperature of $\sim 42$\,K across both lobes, which we used to derive per–channel SiO column densities and molecular masses. The resulting mass, momentum, and kinetic–energy estimates place SMA3 within the regime of heavily mass–loaded molecular jets, while still maintaining the dynamical simplicity and periodicity seen in isolated low–mass systems. The clear chemical stratification between SiO and CS, including the downstream enhancement of CS where the SiO intensity declines, indicates a transition from high–velocity, grain–processing shocks near the jet axis to cooler, denser post–shock regions further along the flow.

Collectively, these results present a coherent, physically self–consistent picture of SMA3 as a compact, magnetically regulated disk–wind system capable of driving fast, chemically rich shocks over several hundred years.

\section*{Acknowledgements}

This paper makes use of the following ALMA data: ADS/JAO.ALMA\#2023.1.01346.S. ALMA is a partnership of ESO (representing its member states), NSF (USA) and NINS (Japan), together with NRC (Canada), NSTC and ASIAA (Taiwan), and KASI (Republic of Korea), in cooperation with the Republic of Chile. The Joint ALMA Observatory is operated by ESO, AUI/NRAO and NAOJ. This work made use of Astropy: \footnote{https://www.astropy.org} a community-developed core Python package and an ecosystem of tools and resources for astronomy \citep{astropy:2013, astropy:2018, astropy:2022}.

The Cologne Database for Molecular Spectroscopy (CDMS) for spectroscopic data was used to derive molecular properties in this paper \citep{CDMS1,CDMS2}

\section*{Data Availability}

The data underlying this article are available in the ALMA Science Archive at https://almascience.eso.org/asax/, under project code 2023.1.01346.S


\bibliographystyle{mnras}
\bibliography{references} 



\appendix

\bsp	
\label{lastpage}
\end{document}

%% file: observations.tex
\begin{table*}
    \centering
    \caption{Observational parameters}
    \label{tab:obs-pars}
    \begin{tabular}{lrlllrrrrr}
        \hline
        Execution & Spectral & Date & Conf.$^\dagger$ & $N_\text{ant}$  & On-source &T$_\textrm{sys}$ &  \multicolumn{2}{c}{Baseline length} & Calibrators\\
        block & setup & & & &time &&min& max& (Flux and\\
        &&&&&(s)&(K) & (m)&(m) & Bandpass)\\
        \hline
        X408f   &225~GHz&2023-10-01 &C-7&45& 845& 92&92&8500 & J0750+1231\\
        X375d   &250~GHz&2023-10-26 &C-7&48& 165& 82&92&8500 & J0423-0120\\
        X7d9d   &250~GHz&2023-10-27 &C-7&48& 917& 85&67&8300 & J0423-0120\\
        X136cb  &250~GHz&2024-01-02 &C-4&46& 212& 94&15&784 & J0423-0120\\
        X28a0   &250~GHz&2024-05-19 &C-4&45& 213& 86&15&784 & J0423-0120\\
        X96e4   &225~GHz&2024-01-01 &C-4&49& 273& 102&15&784 & J0510+1800\\
        X39da   &225~GHz&2024-05-19 &C-4&45& 273&  87&15&784 & J0750+1231\\\hline
        \multicolumn{5}{l}{\small$^\dagger$ Nominal ALMA configuration at time of observation. }\\
    \end{tabular}
    \medskip
    
\end{table*}